Consider avoiding the .05 significance level

David Navon

Yoav Cohen

The University of Haifa, Israel

National Institute for Testing and Evaluation, Israel

*Running Head*: Avoiding .05

*Address*:
David Navon
Department of Psychology
University of Haifa
Haifa 31905
Israel

*Phone*: 972-4-8240927
*E-mail*: dnavon@psy.haifa.ac.il

*Abstract*

It is suggested that some shortcomings of Null Hypothesis Significance Testing (NHST), viewed from the perspective of Bayesian statistics, turn benign once the traditional threshold *p* value of .05 is substituted by a sufficiently smaller value. To illustrate, the posterior probability of $H_0$ stating P=.5, given data that just render it rejected by NHST *with a p value of .05* (and a uniform prior), is shown here to be not much smaller than .50 for most values of N below 100 (and even *exceeds* .50 for N≥100); in contrast, *with a p value of .001* posterior probability does not exceed .06 for N≤100 (neither .25 for N<9000). Yet more interesting, posterior probability becomes quite independent of N with a *p* value of .0001, hence practically satisfying the α *postulate* – set by Cornfield (1966) as the condition for *p* value being a *measure of evidence* in itself. In view of the low prospect that most researchers will soon convert to use Bayesian statistics in any form, we thus suggest that researchers who elect the conservative option of resorting to NHST be encouraged to avoid as much as possible using a *p* value of .05 as a threshold for rejecting $H_0$. The analysis presented here may be used to discuss afresh which level of threshold *p* value seems to be a reasonable, practical substitute.

*Highlights*

- We argue that some shortcomings of NHST are not as irreparable as sometimes presented
- Posterior probability of a H0 rejected with p value of .05 hovers around chance level
- That probability is much smaller and much less dependent on N, the lower p value is
- We suggest using as threshold significance level p values much smaller than .05

Null Hypothesis Significance Testing (NHST) is being employed as the main inference tool in experimental studies despite the longstanding scholarly controversy over its use. That dispute started roughly at the time that Ronald Fisher began criticizing other statisticians, among them J. Neyman and M.G. Kendall, for overstretching and convoluting his rudimentary idea of the null hypothesis, as he had first introduced[1] in Fisher (1925) and brilliantly explained a decade later (Fisher, 1935; cf review in Salsberg, 2001).

The use of NHST grew considerably more controversial in recent decades in light of several concerns (see, e.g., Bakan, 1966; Campbell, 1982; Carver, 1978, 1993; Cohen, 1990, 1994; Edwards, Lindman & Savage, 1963; Falk & Greenbaum, 1995; Hunter, 1997; Krantz, 1999; Krueger, 2001; Lykken, 1968; Nickerson, 2000; Simmons, Nelson & Simonsohn, 2011; Wilson, Miller & Lower, 1967; and lately, Nuzzo, 2014). A prominent concern is that NHST does not measure evidential weight, rather resorts to collapsing a continuum of evidential weight into a binary decision (reject/accept $H_0$), thereby nullifying the potential impact of any piece of evidence that fails to meet, even just barely, the significance criterion, often for quite prosaic reasons (budgetary constraints, for example). Another worry is that NHST focuses on the cost of a false positive while almost neglecting the value of a true positive (Killeen, 2006). Vehement NHST critics, like Armstrong (2007) and Krueger (2001), deplore what one of them (Krueger, ibid, p.16) called "the survival of a flawed method" and others designated as an instance of "trained incapacity" (Ziliak & McCloskey, 2008, p. 238) or of a "cult" (ibid, book's title). Consequently, reliance on confidence intervals and effect sizes has increased in last decades (e.g., Kline, 2004) and is now advocated by quite a few authors to officially replace use of NHST (e.g., Cumming, 2013).

On the other hand, there is no consensus that NHST is basically wrong (see, e.g., Hagen, 1997; Mogie, 2004; Nickerson, 2000). Nickerson chose to summarize his thorough review of the controversy as follows: "NHST is easily misunderstood and misused but when applied with good judgment it can be an effective aid to the interpretation of experimental data" (ibid, p. 241). One way or the other, it is a fact that NHST is still applied to research data almost as prevalently as before (as noted in Lakens & Evers, 2014, p. 284), not the least because it is still required in many publication manuals of scholarly journals.

True, many people doing research in psychology fail to heed the warnings about NHST made in textbooks of statistics for psychologists (and refreshed from time to time in articles such as those cited above). Some of them are not fully aware of its logic, sometimes to the point of failing to notice grave fallacies in its interpretation, such as the fallacy in the claim "$p = .05$ entails that the likelihood that $H_0$ is true is about 1 in 20". For them, at least, a reminder is surely not gratuitous. However, it is doubtful that having been corrected, they would ever become sufficiently disillusioned with NHST as to become amenable to an advice to substitute it with an alternative.

Moreover, the chief alternative to NHST, Bayesian analysis (see, e.g., Kruschke, 2013; Wagenmakers, 2007) - well-founded and elegant as it may be - is unfortunately too sophisticated to be widely used by researchers who have not been sufficiently trained to apply it (and may be questionably satisfactory in itself; see Killeen, 2006, 2006a). Hence, revolutionary ideas, such as a call to ban the use of NHST (Hunter, 1997), seem so far short of being practical for the time being.

A proposal (Wagenmakers, ibid) to adopt a simple shortcut for obtaining posterior probabilities of $H_0$ (the Bayesian information criterion; Bic, for short) is yet to prove sufficiently acceptable and convenient. Using instead Bayes-factor calculators (Morey, Rouder, & Jamil, 2014), albeit possibly more accessible, may not prove very useful until users were sufficiently versed in the logic of Bayesian inference. The fate of another sophisticated methodology recently proposed – the decision theory for science (DTS for short; Killeen, 2006a) - may be found similarly hard to practice for a similar reason.

The repeated exposure of researchers to shortcomings of NHST might eventually amount to a critical mass sufficient for outweighing the resistance and conservatism that as yet keep it afloat. However, granting that it is not clear that a full blown rejection of NHST is on the horizon, perhaps something can be done to attain a more modest, yet feasible, objective that may present the lesser evil <u>within</u> the traditional framework, namely cases in which researchers do apply NHST (or report whether or not a variate value of special theoretical interest - such as µ=0 - falls within a predetermined confidence interval).

Some attempts in that spirit have already been done. Realizing that "NHST is here to stay" (Krueger, ibid, p. 25), some authors presented ways to evaluate evidential value of studies in view of NHST results (Lakens & Evers, 2014), or insisted that hypothesis-testing methods in the broad sense are nonetheless crucial (see Morey, Rouder, Verhagen & Wagenmakers, 2014).

In keeping with that, we suggest that it may be possible, somewhat ironically, to use Bayesian logic to decide how to more reasonably apply NHST, which should allow researchers to keep practicing the latter. Some progress in that direction can be made, once it is acknowledged that although *p* values lie on a continuum, it is possible to re-classify the specific effects of using any particular value as "dubious" and "benign" in a way that is not arbitrary. Moreover, the dividing line differs from the traditional threshold of .05 (or equivalently, a 95% confidence interval).

To show that, let us confine the discussion to the use of NHST and examine the issue by three criteria – (a) posterior probability of $H_0$, (b) dependence of that posterior probability on sample size, (c) replicability.

*Posterior probability and its dependence on sample size*

For a start, consider the first two criteria. In the following, we extend an argument made by Wagenmakers (2007) to demonstrate a grave drawback of NHST warranting conversion to Bayesian inference, to rather suggest a way to counter that drawback within the framework of NHST in a quite simple way.

The drawback pointed to by Wagenmakers is that a *p* value does not qualify as a measure of statistical evidence, since it fails to satisfy the requirement (denoted by him as the *p postulate*; cf the term α *postulate* in Cornfield, 1966) that "identical *p* values convey identical levels of evidence, irrespective of sample size" (Wagenmakers, 2007, pp 779-780). He illustrated that failure by showing that the posterior probability of $H_0$ (e.g., one stating that the binomial parameter Θ, specifying the proportion of opting for a specific alternative in a set of choices between given two alternatives, equals .5), given data having *p* = .05, is quite high - falling just below .50 for *N*s smaller than about 60, and furthermore grows dramatically as *N* increases - exceeding .50 approximately with *N*≥100.

Wagenmakers is right of course. Yet, it should be noted that although *p* values do not in principle satisfy the *p* postulate, *in practice* the extent by which the *p* postulate is violated by NHST may be reasonably tolerable with small enough *p* values.

To inspect that, we applied Wagenmakers' procedure[2] to a few other customarily used significance levels (.01, .001, .0001) in addition to .05 (Navon & Cohen, 2008). We used only the most conservative prior - the objective Beta (1,1), which is actually a uniform distribution over the whole range, and selected levels of *N* (20, 40, 60, 80, 100, 200, 400, 600, 800, 1000, 2000, 4000, 6000, 8000, 10000), calculations for each are shown as separate points on the plotted graphs.

The results are presented in Figure 1A. The slopes of the graphs relating posterior probability to N (within the range of commonly used N levels) become progressively smaller the lower the *p* value is: The slope that is large with a *p* value of .05 is still about as large with a *p* value of .01, but is considerably moderated with a p value of .001. With a *p* value of .0001, the posterior probability of $H_0$ becomes almost independent of *N*, at least for most practical purposes (also, note the corresponding graph in Figure 1B, which is a magnified view of the leftmost side of Figure 1A).

The growing robustness to *N* with more and more extreme *p* values imparts in itself increasing confidence that the decision to reject $H_0$ is founded on a firm basis. Yet, *confidence* apart, how does level of significance affect *evidential import*?

Unlike cases where the *p* value is .05 (and for that matter, also with 95% confidence intervals), with *p* values that are more extreme, the posterior probability of $H_0$ is most often comfortably smaller than .50: With a *p* value of .0001, the probability is smaller than .06 throughout the range. With a *p* value of .001, the probability is smaller than .10 up to an *N* of about 1000, and reaches .25 only with an *N* of about 10000. With a *p* value of .01, the probability is smaller than .50 up to an *N* of about 1250. Consider, for example *N*=100. As can be seen in Figure 1B, whereas the posterior probability is about .52 with a *p* value of .05, it is about .22, .08, or less than .01 with *p* values of .01, .001., or .0001, respectively.

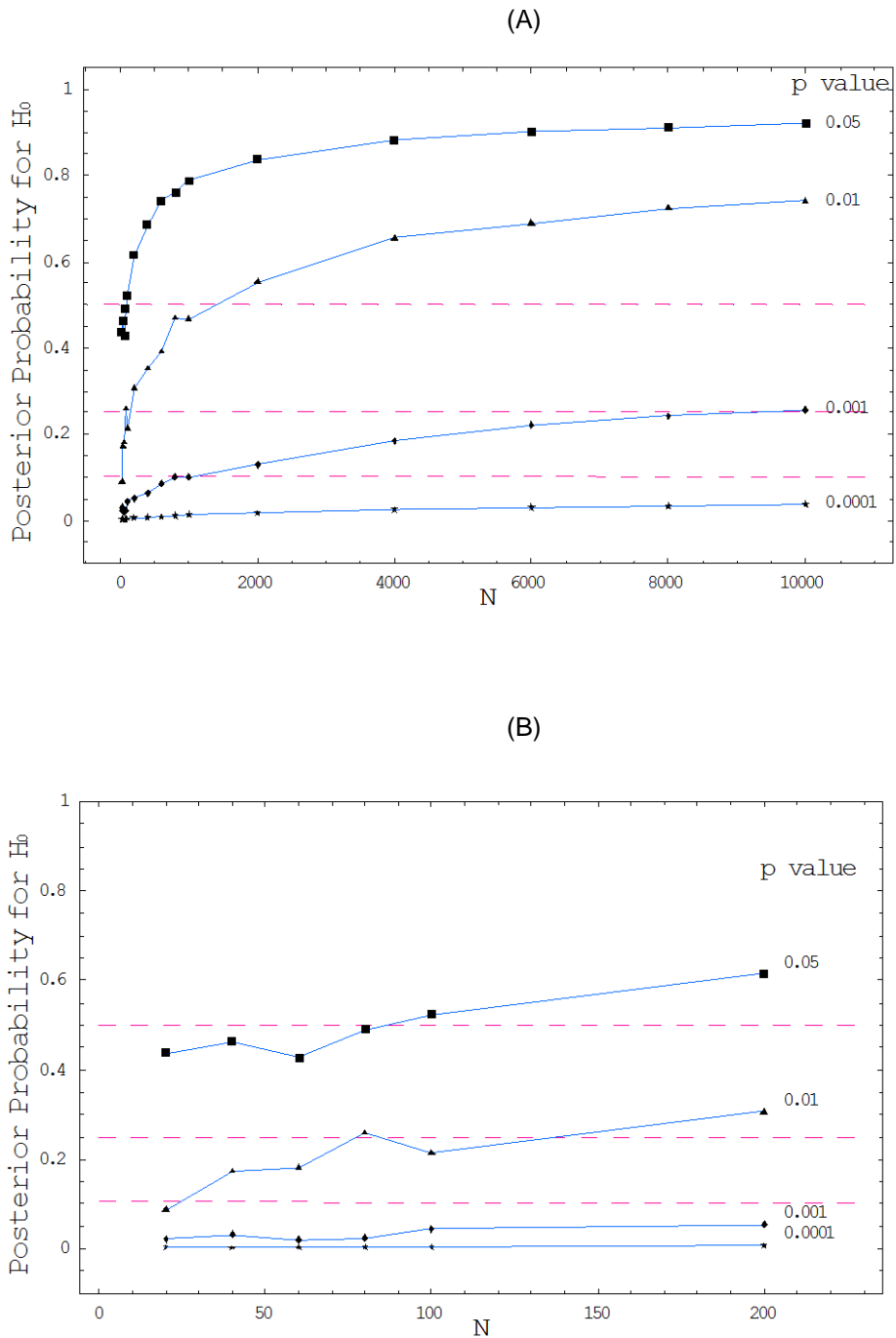

*Figure 1.* The posterior probability of $H_0$ (stating that the Binomial parameter $\Theta$ equals .5) [given (a) data that just render it rejected by NHST, (b) the objective prior], as a function of *N* (ranging from 20 to 10000 in increments specified in the text in panel A; ranging from 20 to 200 in panel B) and *p* value (05, .01, .001, or .0001).

*Notes*: (a) The dashed horizontal lines (plotted at ordinate values of .10, .25 and .50) are meant just for reference. (b) The jitter apparent at some places along the graphs is due to the fact that since the variable "number of successes" is discrete, the *p* value corresponding to any given significance level often only approximates it; when the gap in actual *p* values for successive *N*s is considerable, which might occur with small *N*s, the monotonicity of the $\Pr(H_0)$-*N* graph may be violated.

Thus, the results for *p* values more extreme than .05 certainly look far more benign than the results for *p* = .05. The implication is that researchers are quite justified in concluding that the posterior odds of $H_0$ (given the data of just a single experiment and a uniform prior) is clearly < 1, most often < 1/3, whenever having obtained a rejection with *p* ≤ .001, let alone *p* ≤ .0001. They are also right in concluding that the odds are < 1/2, and most often < 1/3, whenever having obtained a rejection with *p* ≤ .01 given that *N* ≤ 200 (see Figure 1B).

The latter condition is very frequently met in empirical studies. To illustrate (without any pretense for representative sampling): Out of the 42 experiments reported in some given issue of *Psychonomic Bulletin and Review* (Vol. 14, no. 5, October 2007), <u>all</u> used *N*s < 200 (furthermore, 88% used *N*s < 100 and 64% used *N*s < 50). A further illustration concerns just tests of $H_0$ stating P=.5: Out of the first 32 experiments found in a Google Scholar search with keywords "psychology" and "sign test", <u>all</u> used *N*s < 200 (furthermore, 84% used *N*s < 100 and 59% used *N*s < 50).

All that seems to call for further evaluation of the time-honored tradition to consider .05 as the threshold *p* value allowing rejection of the null hypothesis. Note, a rejection with *p* ≤ .05 only enables one to determine that the posterior probability of $H_0$ is < .50, namely that the posterior odds are smaller than 1, with *N*s not greater than about 80. One may argue that that is an accomplishment decent enough for a single experiment given a uniform prior, actually about all a realistic researcher may ask for, considering that the scientific community rarely passes judgment on the basis of only a single experiment (see Krantz, 1999; Nickerson, 2000). On the other hand, the analysis above casts some doubts on the evidential import of a result with a *p* value of .05. To the extent that the doubts are grave enough to worry about, the analysis at the same time suggests that there is a straightforward remedy for that – adopting a more stringent threshold *p* value.

The traditional use of .05 as a threshold *p* value, namely the practice of reporting a result as significant only once its *p* value is at least as small as .05, must be somewhat arbitrary[3]. Any alternative way to set a threshold must be somewhat arbitrary as well. However, in retrospect one may prefer that the adopted threshold would severely reduce the number of cases in which the posterior odds of $H_0$, given the most conservative prior, would hover around 1. Resetting the threshold to .01, the second most customary level, yields odds of about 1/3 or even less for the most commonly used *N* values. That seems a bit less arbitrary. Resetting it to .001 yields odds of about 1/20 or even less for those values of N. Resetting it to .0001 yields much smaller odds yet, hovering somewhere near 1/100.

It is not for us to tell what the threshold *p* value should be, neither to recommend how that should be institutionally decided. Suffice it to note that it is possible to reset it in a way that would render the posterior odds minute enough, and its dependence on *N* small enough, to satisfy even the most stringent. For a reasonable idea about how to do that, note suggestions made by Johnson in another article (Johnson, 2013), presenting conclusions roughly similar to those we independently make here (some of which already mentioned in Navon & Cohen, 2008).

*Replicability*

Another, not unrelated, reason to consider this suggestion is the substantial difference between .05 and .01 (let alone yet-more-extreme *p* values) vis-à-vis another criterion – replicability - argued by some to be no less meaningful as an inference tool than *p* values. Following Greenwald, Gonzales, Harris & Guthrie (1996) who defined replicability as the probability of rejecting again (with a *p* value of .05 at most) a false null hypothesis, Krueger (2001) expanded the definition for all possible values of prior probability of $H_0$. He illustrated, for three selected values of $Pr(H_0)$, how replicability depends on the value observed in the first rejection (see ibid, Figure 2).

In a recent large-scale open science collaboration (Aarts et al, 2015) meant to assess rates of replication (using the definition suggested by Greenwald et al, ibid) of 100 studies published in all 2008 issues of three psychology journals, it was found that whereas only 18% original studies with *p*<0.04 were replicated, replication rate was 63% for original studies with p<0.001.

Killeen (2005) developed a general formula for transforming a *p* value to the probability of replicating the *sign* of a statistically significant effect that he termed $p_{rep}$ (cf. Killeen, 2005, 2006; Sanbaria & Killeen, 2007). That proposed alternative to NHST received quite a few interpretations as well as critical reactions (see short review of commentaries in Cumming, 2010). Later, Miller & Schwarz (2011) raised further doubts about the potential of the $p_{rep}$ measure to *supplant* NHST, due to what they considered "formidable source of uncertainty associated with estimation of replication probability for a particular experimental effect of interest" (ibid, p. 359).

Yet, that measure, inadequate in itself as it may be qua *substitute* for NHST, may at least inform us how to practice NHST, by providing some gross idea of how the notion of replicability (however being specifically formulated) *relates to p value*, for example whether an effect found to have a *p* value of .05 could be satisfactorily replicable.

For whatever it is worth to gauge replicability, the $p_{rep}$ measure indicates that replicability heavily depends on *p* value: According to that index, whereas replicability of a result having a *p* value of .05 is ~.877, replicability is considerably higher for results significant with a *p* value of .01 (~.950), let alone with a *p* value of .001 (~.986), much less .0001 (~.996). In other words, while a result having a *p* value just barely significant at the .05 level stands a considerable (.123) chance of failing to replicate, results having *p* values of about .01, .001 or .0001 are considerably less likely (.050, .014, or .004, respectively) to do that.

Since it seems quite prudent to be worried about reporting a result as significant in case the replicability is not high enough (cf Greenwald et al, 1996 or Nuzzo, 2014), it makes sense to reset the threshold significance level (at least for a first rejection[4]) to .01, .001 or .0001, as deemed sufficiently satisfactory..

*Summary*

The two arguments discussed above suggest that resetting the threshold *p* value to .01, or even .001 or .0001, is not arbitrary at all. It is quite reasonable to expect that implementing that change would reduce *both* the rate of rejection of null hypotheses that are likely to be true (in the light of their posterior probability) and the rate of failures to replicate a result reported as significant.

Obviously, the cost of such a move would be the need to increase power of tests, typically by collecting considerably more data (e.g., running more lab animals, recruiting more human respondents). Yet, as can be seen in Figure 1B, unlike cases in which power is increased to attain rejection of $H_0$ at the .05 level - which results in *reduced* diagnosticity (viz, reduction in posterior odds in favor of $H_1$; or put more plainly, decrement in the evidential value that $H_0$ is false) - power increase invested in making $H_0$ rejected at the .01 level (rather than at the .05 level), let alone the other two levels, helps to gain a substantial *rise* in diagnosticity.

For example, when testing $H_0: \mu=40$ against a true alternative $\mu=43$, given $\sigma=8$, in order to obtain a $\beta$ level of at most .02, minimal N should be 98 with $\alpha=.05$, yet 137 with $\alpha=.01$. As can be seen in Figure 1B, that modest increase (of ~40%) in N yields a quite substantial rise in diagnosticity - from ~1 to ~3 (namely, by ~200%). Returns of that order of magnitude would seem to warrant the added resource costs. That seems worthy of doing despite anticipated reluctance of researchers to undertake the extra burden.

The question is what it would take to bring about such a reform. First, a decent support has to be garnered within the academic community. Subsequently, a change in behavior should follow.

It is not clear how that could best be accomplished. We are not advocating anything in particular, certainly not enforcement through an intra-disciplinary institutional legislation. The gamut of possible alternatives to that is quite broad.

An effective initial step would be the revision of textbooks. Students who are taught in undergraduate studies that an F statistics is considered significant only if its *p* value is no larger than .01, or .001 or .0001 (along with an historical note that people once used to pass as significant even *p* values as large as .05) are likely to internalize that and follow suit. Another step would be to bring about change in editorial policy and attitudes. Not long after journal editors start exhibiting discomfort towards results with *p* values less extreme than the reset threshold level, authors will presumably start adjusting their subjective standards of suitability for journal submission correspondingly.

*Footnotes*

1.  Yet, while stressing that "tests of significance… should not be confused with automatic acceptance tests" (Fisher, 1958, p. 128), Fisher was far from questioning the logic of NHST, as he wrote in his preface to the 13<sup>th</sup> edition of *Statistical methods for research workers*: "Today exact tests of significance need no apology" (ibid, p. *v*).

2.  Which amounts to the following, using Wagenmakers's own words (ibid, p. 792), with a few omissions and slight paraphrasing:
    > The number of observations $N$ was varied… and for each of these $N$s we determined the number of successful decisions $s$ that would result in an NHST $p$ value that is barely significant at the .05 level, so that for these data $p$ is effectively fixed at .05… Next, for the data that were constructed to have the same $p$ value of .05, we calculated the Bayes factor $Pr(D|H_0) / Pr(D|H_1)$ using the objective Beta (1,1) prior. From the Bayes factor, we then computed posterior probabilities for the null hypothesis.

    Specifically, the posterior probability of the null hypothesis, $Pr(H_0)$, equals $l_0 / (l_0 + l_1)$, where $l_0$ is the likelihood of the observed number of successes in $N$ trials given $\Theta = .5$ (as the point hypothesis $H_0$ has it), and $l_1$ is the integral of the likelihoods (of the observed number of successes in $N$ trials) across all values of $\Theta$ within the interval [0,1] (as the interval hypothesis $H_1$ has it).

3.  To illustrate, in an early reference to the notion of significant deviation (probably as early as the first edition of *Statistical methods for research workers*, published in 1925) Fisher himself mentioned the .05 level as just a *conceivable* threshold, as implied by the "if" in "if we take P=.05 as the limit of significant deviation…" (p. 82 in the 13<sup>th</sup> edition).

4.  Once a null hypothesis has been rejected with reasonable confidence, so that - to use Bayesian terminology - $Pr(H_0)$ is considerably smaller than .5, there is less ground to worry about $p$ values larger than .01. Actually, even a $p$ value of .10 or .15 may be considered as a replication, albeit a weak one. If and when a meta-analysis is performed, all the recorded data would be input conjunctively and have their impacts respective with each one's diagnostic value, high or low as the case may be.